\documentclass[letter]{ieice}
\usepackage{graphicx,xcolor}
\usepackage[fleqn]{amsmath}
\usepackage{newtxtext}
\usepackage[caption=false]{subfig}
\usepackage{url}

\field{D}
\title{Disaggregated Accelerator Management System for Cloud Data Centers}
\titlenote{This paper extends our preliminary work published at
SC2018 research poster~\cite{Takano2018.SC}. Specifically, we conducted additional
experiments with more realistic deep learning workload.
Moreover, we have clalified the novelty in related work section.}

\authorlist{%
 \authorentry[takano-ryousei@aist.go.jp]{Ryousei Takano}{m}{AIST}\MembershipNumber{1617664}
 \authorentry{Kuniyasu Suzaki}{m}{AIST}\MembershipNumber{8912341}
}
\affiliate[AIST]{The author is with Information Technology Research Institute,
National Institute of Advanced Industrial Science and Technology (AIST), Tsukuba, 305-8560 Japan.}

\received{2020}{1}{1}
\revised{2020}{1}{1}

\begin{document}
\maketitle
\begin{summary}
A conventional data center that consists of monolithic-servers is confronted with
limitations including lack of operational flexibility, low resource utilization,
low maintainability, etc. Resource disaggregation is a promising solution to
address the above issues. We propose a concept of disaggregated cloud data center
architecture called Flow-in-Cloud (FiC) that enables an existing cluster computer
system to expand an accelerator pool through a high-speed network.
FlowOS-RM manages the entire pool resources, and deploys a user job on
a dynamically constructed slice according to a user request.
This slice consists of compute nodes and accelerators where each accelerator is
attached to the corresponding compute node.
This paper demonstrates the feasibility of FiC in a proof of concept experiment
running a distributed deep learning application on the prototype system.
The result successfully warrants the applicability of the proposed system.
\end{summary}

\section{Introduction}

The end of Moore's law is coming within a decade due to technical and economic limitations.
No more drastic performance improvement for general purpose processors is expected and
new computing paradigms and architectures are needed for the explosive growing computational
workload such as big data analysis, deep learning training and inference, and so on.
Specialization or, in other words, domain specific architecture~(DSA) is a promising research
direction in the Post-Moore era.
Specifically, many task-specific accelerators including Google TPU, Fujitsu DLU,
Microsoft BrainWave, and D-Wave Quantum Annealer were proposed recently.
To take advantage of such accelerators, it is important to establish a resource management
system to fully utilize a variety of hardware resources, including a generic processor,
an accelerator, and storage, depending on the workloads.
However, a conventional data center consists of monolithic servers and it cannot provide
such flexible use of computing hardware resources.
It also faces limitations including lack of operational flexibility, low resource utilization,
low maintainability, etc~\cite{Gu2017.NSDI,Shan2018.OSDI}.

To address the limitations of conventional data centers, there is an emerging interest in
resource disaggregation~\cite{Asanovic2014.FAST,Katrinis2016.DATE,Guo2019.OFC,Gu2017.NSDI,Shan2018.OSDI}
that decomposes monolithic servers into independent hardware components, including CPU,
accelerator, memory, and storage, through a high-speed network.
In a disaggregated data center, hardware components are separated in each resource pool
and reconstructed to meet the user requirements.
We have proposed a new concept of disaggregated data center architecture, Flow-in-Cloud~(FiC),
that enables an existing PC cluster to expand an accelerator pool through a high-speed
network. FlowOS manages hardware resources and application jobs on FiC.
To demonstrate the feasibility of FiC and FlowOS, currently we are developing FlowOS on
the prototype system of FiC.
This paper focuses on the resource management system of FlowOS called FlowOS-RM and
reports on the effectiveness for distributed deep learning applications.

\section{Related Work}\label{sec:related}

Some papers~\cite{Gu2017.NSDI,Shan2018.OSDI} have reported that the resource utilization,
e.g., CPU or main memory, varies considerably for each server in commercial data centers.
This is because it is quite difficult to assign various workloads in such a way that
all resources are fully and equally consumed. As a result, the resource utilization
remains low.
To address this problem, many studies of resource disaggregation have emerged since
the middle of the 2010s. This movement started with hardware-level resource disaggregation
~\cite{HPE.TheMachine,Asanovic2014.FAST,Katrinis2016.DATE,Guo2019.OFC}
and expanded to the OS-level resource disaggregation~\cite{Gu2017.NSDI,Shan2018.OSDI}
in recent years.
The interconnection technology for enabling disaggregation, including Gen-Z, is being
standardized and will be commercialized in the near future.
The proposed system addresses device disaggregation with a special focus on accelerators,
and it seamlessly extends an existing cloud data center by facilitating access an accelerator resource pool.
This characteristic allows us to benefit from the software ecosystem of an existing
cluster resource management system like Apache Mesos.
To the best of our knowledge, there is no existing works that consider cooperation with
cluster resource management systems.
Although we use an electric interconnection network,
an optical network is promising as several researchers have proposed in
\cite{HPE.TheMachine,Asanovic2014.FAST,Katrinis2016.DATE,Guo2019.OFC}.

Several device disaggregation technologies have been proposed in
~\cite{Suzuki2006.HOTI,Duato2010.HPCS}.
ExpEther~\cite{Suzuki2006.HOTI} is a PCIe-over-Ethernet technology and it allows us to
dynamically attach and detach remote PCIe devices through Ethernet.
On the other hand, rCUDA~\cite{Duato2010.HPCS} is an OS-level disaggregation technology.
Although it works with only NVIDIA GPUs, it can seamlessly access a remote GPU through
Infiniband and Ethernet.
Our preliminary experiment~\cite{Takano2018e.SIGHPC} shows
the performance overhead of device disaggregation technologies. The host-to-device
bandwidth of ExpEther is about 20\% of that of a local PCIe device.
While this performance degradation is a worse-case situation, some application
performances are regulated by the amount of traffic among
the host and devices. On the other hand, the impact on computation bound applications
like GEMM and convolution is negligible.
The practical problem with rCUDA is the lack of compatibility.
For example, it does not support cuDNN, which is heavily used on deep learning applications.

\section{Flow-in-Cloud}\label{sec:FiC} 

\begin{figure}[t]
 \begin{center}
 \includegraphics[width=\linewidth]{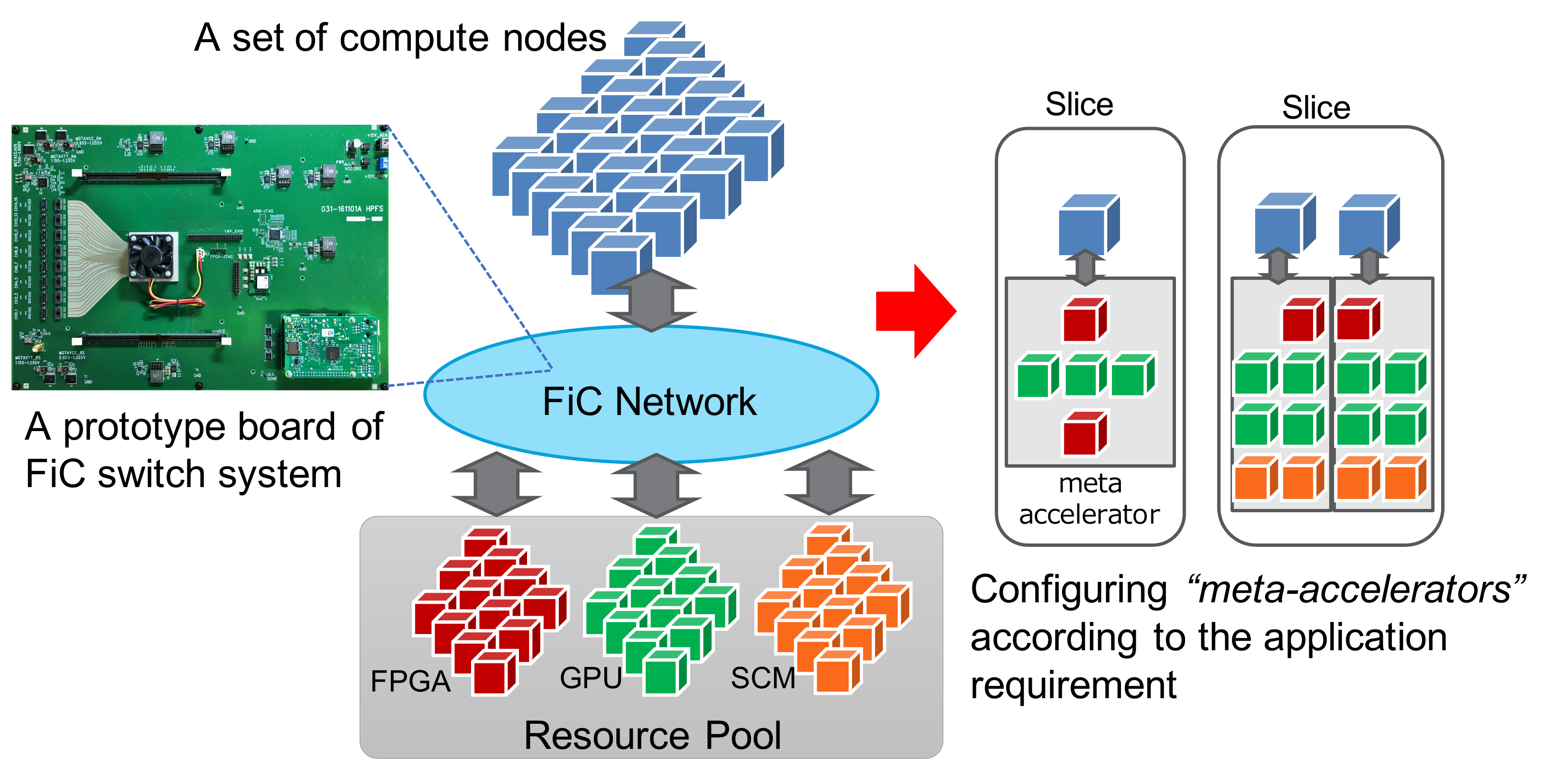}
 \end{center}
 \caption{The overview of Flow-in-Cloud Architecture}
 \label{fig:FiC}
\end{figure}

Flow-in-Cloud (FiC) is a proof of concept system for a disaggregated cloud data center,
and provides the user with a slice of resources as an application execution environment.
An application is then divided into several tasks and each task is optimized through the use of
a suitable accelerator. In the case of deep learning, a convolution layer task is
executed on GPU, and a full connected layer task is executed on FPGA.
We call such a set of accelerators as a meta accelerator.
A slice is dynamically configured by combining meta accelerators and attaching them
to corresponding compute nodes according to a user request, as shown in Figure~\ref{fig:FiC}.
Compute nodes and accelerators are connected through a high-speed circuit-switched
network called FiC network, and it comprises a set of FiC switch boards~\cite{Hironaka2019.SNPD}
that have a middle grade FPGA chip (Xilinx Kintex Ultrascale XCKU095), 32-10Gbps FireFly
serial connections and DRAM. In addition, Raspberry Pi 3 is implemented on the board
as a controller. Raspberry Pi 3 communicates with FPGA via GPIO for configuration of
FPGA and data transmission.
Note that we employ circuit switching instead of packet switching because
friction-less transition from electric network to optical network is possible.
A circuit-switching logic and a user-defined logic written in a high-level
synthesis language are running on the FPGA, and the latter logic is partially
reconfigurable in advance of application deployment.

FlowOS manages the entire pool of FiC resources, and supports the execution of
a user job on provided slices. FlowOS employs a layered architecture including
FlowOS-Job, FlowOS-RM, and FlowOS-drivers. FlowOS-Job is a heterogeneous programming
framework that allows the users to describe a job composed of several tasks, where
each task is optimized for a specific accelerator according to the workload.
FlowOS-RM is a resource manager and the detail is described in Section~\ref{sec:FlowOS-RM}.
FlowOS-driver is a proxy component to access underlying hardware resources
like accelerator pools and compute nodes.

Currently, we have implemented FlowOS on a small disaggregated data center environment
where compute nodes and accelerators are connected through ExpEther~\cite{Suzuki2006.HOTI}
instead of FiC network.
FlowOS provides two major features: disaggregated resource management and heterogeneous
programming framework as mentioned above. In this paper, we focus on the latter and
demonstrate disaggregate device management and cooperation with a cluster
resource management system.
Although ExpEther cannot support direct communication among accelerators as FiC originally addresses,
it is a reasonable alternative technology to demonstrate the concept of FiC using
commodity hardware.

\section{FlowOS-RM}\label{sec:FlowOS-RM}

FlowOS-RM seamlessly works in cooperation with a cluster resource management
system such as Apache Mesos~\cite{Hindman2011.NSDI}, Kubernetes, SLURM, and so on.
In other words, FlowOS-RM extends such systems to support accelerator disaggregation.

FlowOS-RM works in cooperation with the following components:
(1) Disaggregate device management: ExpEther is a PCIe-over-Ethernet technology
and it allows us to dynamically attach and detach remote PCIe devices through Ethernet.
(2) OS deployment: Bare-Metal Container (BMC)~\cite{suzaki2016.HPCC} constructs
an execution environment to run a Docker image with an application optimized OS
kernel on a node.
(3) Task scheduling and execution: FlowOS-RM is implemented on top of a Mesos
framework, and it co-allocates nodes to meet a user requirement and launches
a task on each node in the manner of Mesos.

\begin{figure}[t!]
 \begin{center}
 \includegraphics[width=\linewidth]{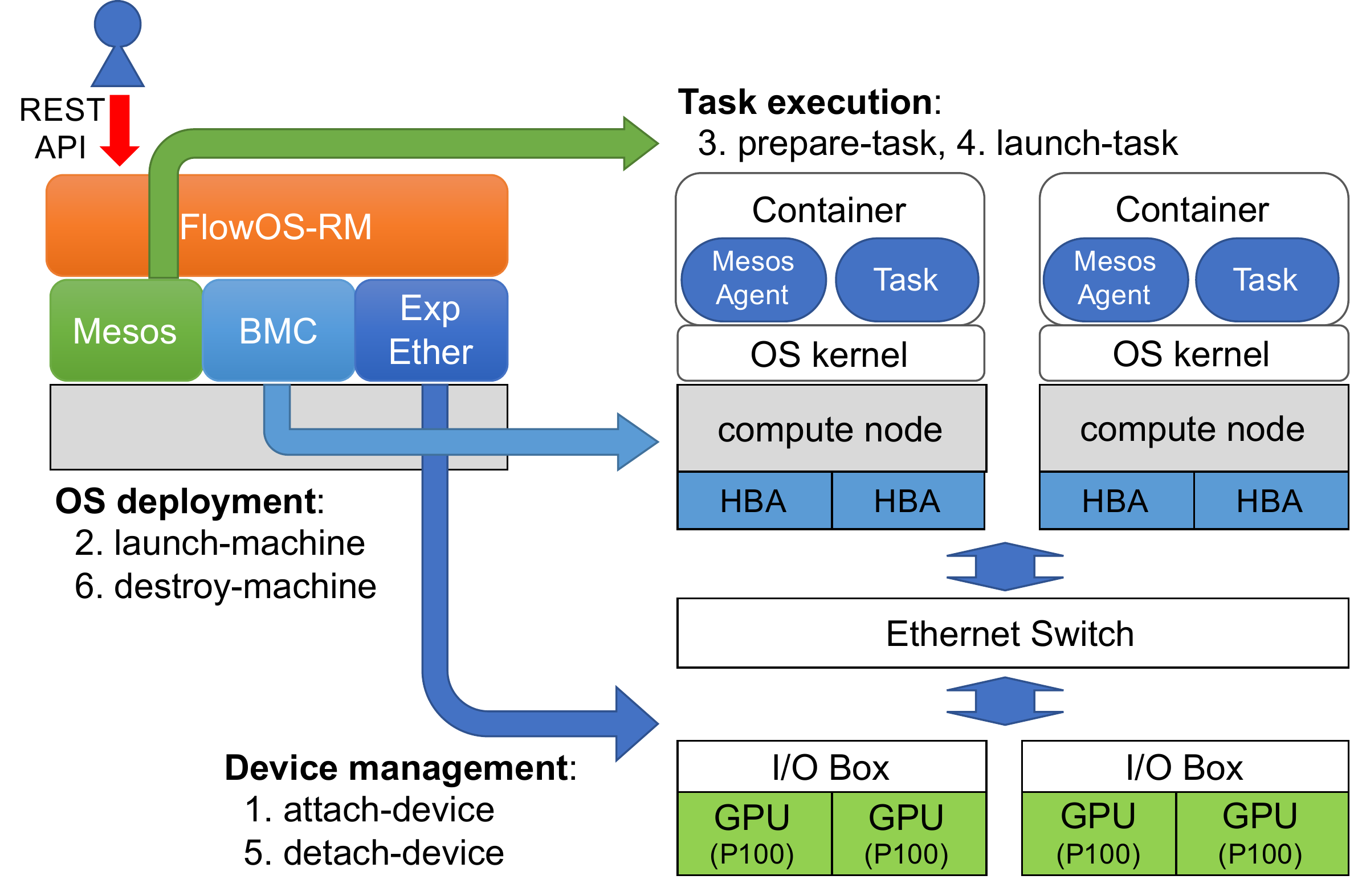}
 \end{center}
 \caption{Job execution flow in FlowOS-RM}
 \label{fig:FlowOS}
\end{figure}

FlowOS-RM provides users with the REST API to configure a slice and execute a job
on it.
A single-node job as well as an MPI type multi-node job is supported.
Figure~\ref{fig:FlowOS} presents a job execution flow in FlowOS-RM, where a job
is a set of tasks and each task runs on a node belonging to a slice.
Table~\ref{tbl:FlowOS} summarizes each operation of FlowOS-RM.
First, a slice is constructed in two steps: \textit{attach-device} and
\textit{launch-machine}. Second, a job is launched in the following
two steps: \textit{prepare-task} and \textit{launch-task}.
After job execution, the slice is destructed in two steps:
\textit{detach-device} and \textit{destroy-machine}.

\begin{table}[t!]
\caption{Major operations in FlowOS-RM}
\label{tbl:FlowOS}
\begin{center}
\begin{tabular}{l|p{5cm}}
\hline \hline
attach-device & Attach devices to a node\\
launch-machine & Boot a node with a specific OS kernel and container, and
              it joins active nodes under Mesos\\
prepare-task & Do housekeeping for launching a task, including submitting
              a task to the corresponding node through Mesos\\
launch-task & Launch a task in a node (running state)\\
detach-device & Detach devices from a node\\
destroy-machine & Shutdown a node and it leaves from active nodes\\
\hline
\end{tabular}
\end{center}
\end{table}

\section{Experiment}\label{sec:expr}

\subsection{Experimental Setting}

\begin{figure}[t!]
  \centering
  \subfloat[Physical Cluster Configuration\label{fig:expr1}]{\includegraphics[width=.8\linewidth]{%
    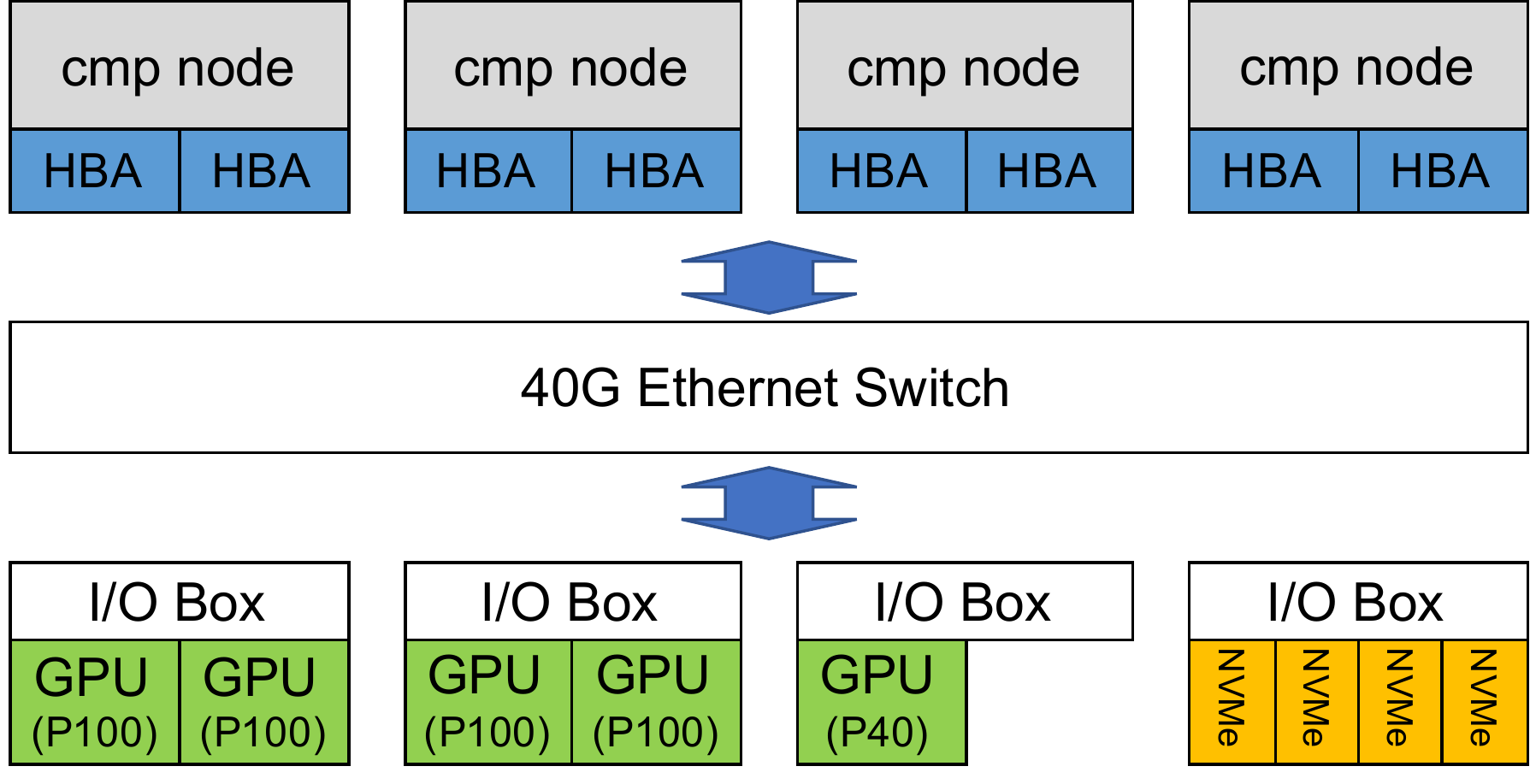}} \qquad
  \subfloat[Slice Configurations\label{fig:expr2}]{\includegraphics[width=\linewidth]{%
    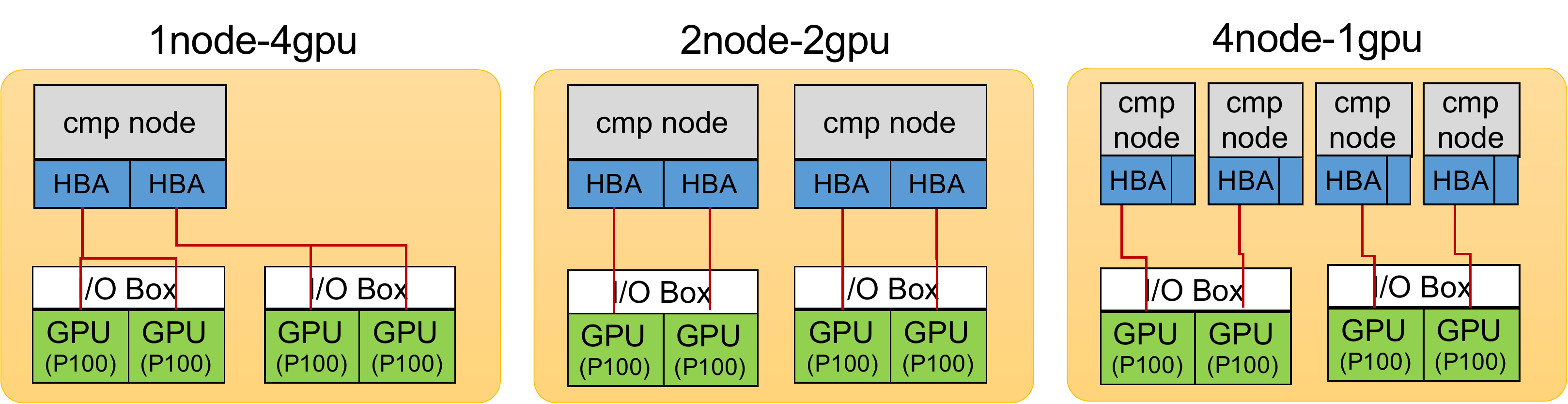}}
  \caption{Experimental Configuration}
  \label{fig:expr}
\end{figure}

In order to demonstrate the feasibility of FlowOS-RM, we have conducted distributed
deep learning training experiments on a four-node cluster environment as shown in
Figure~\ref{fig:expr1}.
Each compute node has two ExpEther HBAs to connect PCIe devices, e.g., NVIDIA Tesla
P100 GPU and Intel NVMe SSD, on I/O Boxes through a 40 GbE swtich.
Linux is running on each compute node, and FlowOS-RM and Mesos are installed on
this environment.
We used two applications, a handwriting character recognition (MNIST) and a large-scale
image classification (ImageNet), as benchmark programs, and they are implemented with
a distributed deep learning framework ChainerMN~\cite{Akiba2017.MLSys}.

\begin{table}[!t]
\caption{Experimental Setting}
\label{tbl:expr}
\centering
\begin{tabular}{l|l}
\hline \hline
\multicolumn{2}{c}{Compute Node Configuration} \\
\hline
CPU & 10-core Intel Xeon E5-2630v4/2.2GHz \\
M/B & Supermicro X10SRG-F \\
Memory & 128~GB DDR4-2133 \\
Network & ExpEther 40G HBA \\
NIC & Intel I350 (Gigabit Ethernet) \\
\hline
\multicolumn{2}{c}{Disaggregated Resource (PCIe device)} \\
\hline
GPU & NVIDIA Tesla P100 x4, P40 x1 \\
NVMe & Intel SSD 750 x4 \\
\hline
\multicolumn{2}{c}{Software Configuration} \\
\hline
OS & CentOS 7.4 \\
Kernel & Linux 3.10.0-514.26.2.el7.x86\_64 \\
& Mesos 1.4.1, ChainerMN, OpenMPI 3.1.0 \\
& NVIDIA CUDA 8.0.61 \\
\hline
\end{tabular}
\end{table}

\subsection{Experimental Results}
\subsubsection{Slice construction and destruction overheads}
\begin{figure}[t!]
  \centering
  \subfloat[MNIST on three slice configurations\label{fig:mnist}]{%
    \includegraphics[width=.9\linewidth]{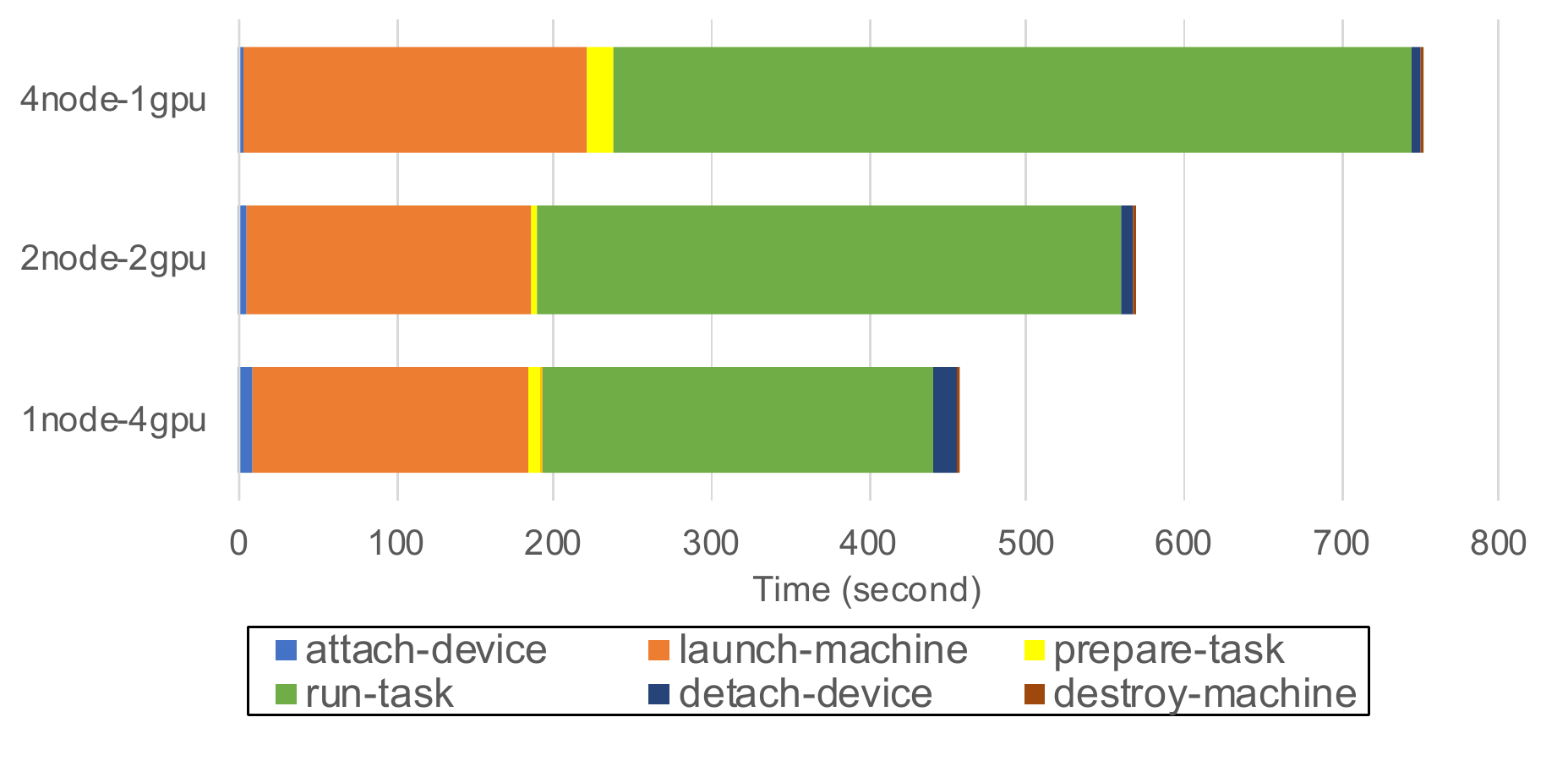}} \qquad
  \subfloat[ImageNet on 4node-1gpu slice configuration\label{fig:imagenet-4node}]{%
    \includegraphics[width=\linewidth]{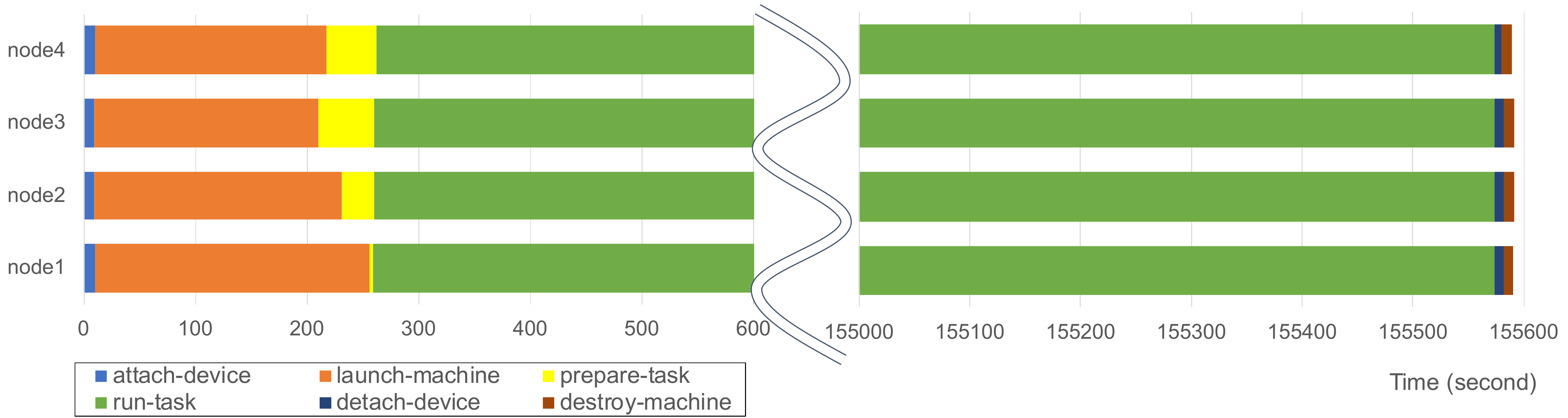}} \qquad
  \subfloat[ImageNet in 2node-2gpu slice configuration\label{fig:imagenet-2node}]{%
    \includegraphics[width=\linewidth]{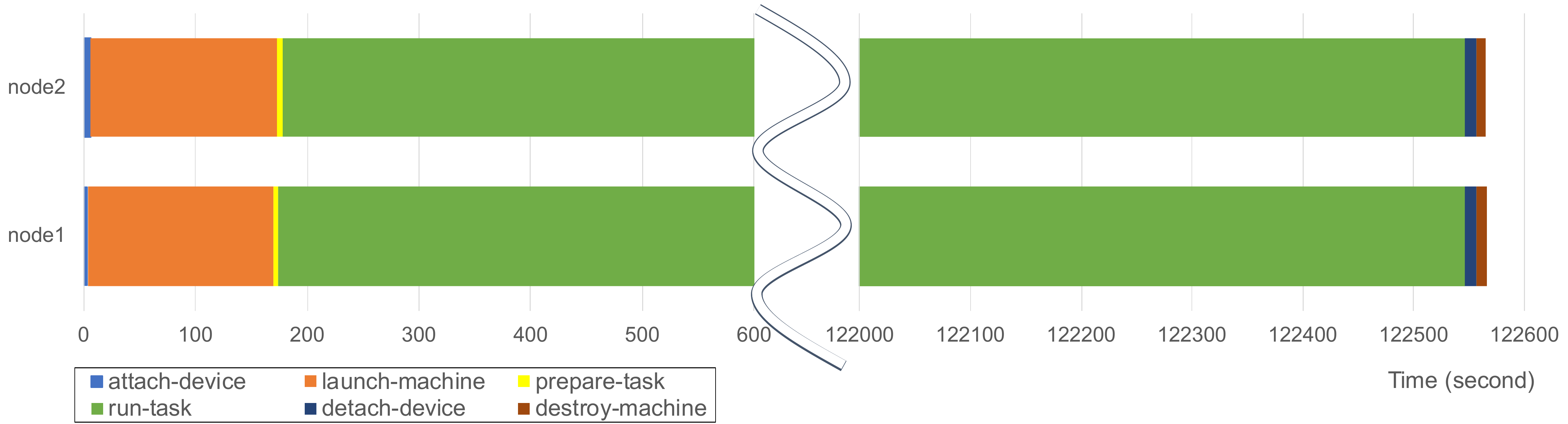}} \qquad
  \caption{Slice execution life cycle}
  \label{fig:expr.exectime}
\end{figure}

We have demonstrated a flexible resource management and the performance overhead
of FlowOS-RM.
Firstly, we ran an MNIST application on three different slice configurations
as shown in Figure~\ref{fig:expr2}, and the breakdown in the execution time for
each slice is shown in Figure~\ref{fig:mnist}.
An MNIST training runs faster as the number of GPUs per node increases.
The \textit{run-task} elapsed times of 4node-1gpu, 2node-2gpu, and 1node-4gpu are
366.36, 237.31, and 104.57 seconds, respectively.
It is a relatively lightweight workload and the slice construction
and destruction operations account for 32\% to 45\% of the total execution time.
Specifically, a \textit{launch-machine} operation takes longer as the number of
nodes increases, because downloading a container image that is about 3GB in size
through GbE becomes the bottleneck.
Some operations including \textit{attach/detach-device} and \textit{launch-task}
take longer as the number of GPUs per node increases, because these operations
are not parallelized.
We plan to reduce the above overhead by using a faster network and parallelizing
operations.

Secondly, we ran ImageNet, a more practical application on the same slice configurations.
We used the ResNet-50 model and ILSVRC2012 dataset. In this experiment, a slice has not only
GPUs but also two NVMe SSDs to store ILSVRC2012 dataset.
Unlike an MNIST experiment, the slice construction and destruction operations account
for 0.15\% to 0.17\% of the total execution time as shown in Figures~\ref{fig:imagenet-2node}
and \ref{fig:imagenet-4node}.
Generally speaking, deep learning training execution time tends to significantly increase and
the overhead of FlowOS-RM can be negligible.

\subsubsection{Resource sharing}
\begin{figure}[t!]
  \centering
  \subfloat[Slice Configurations\label{fig:mnist-mix}]{\includegraphics[width=.8\linewidth]{%
    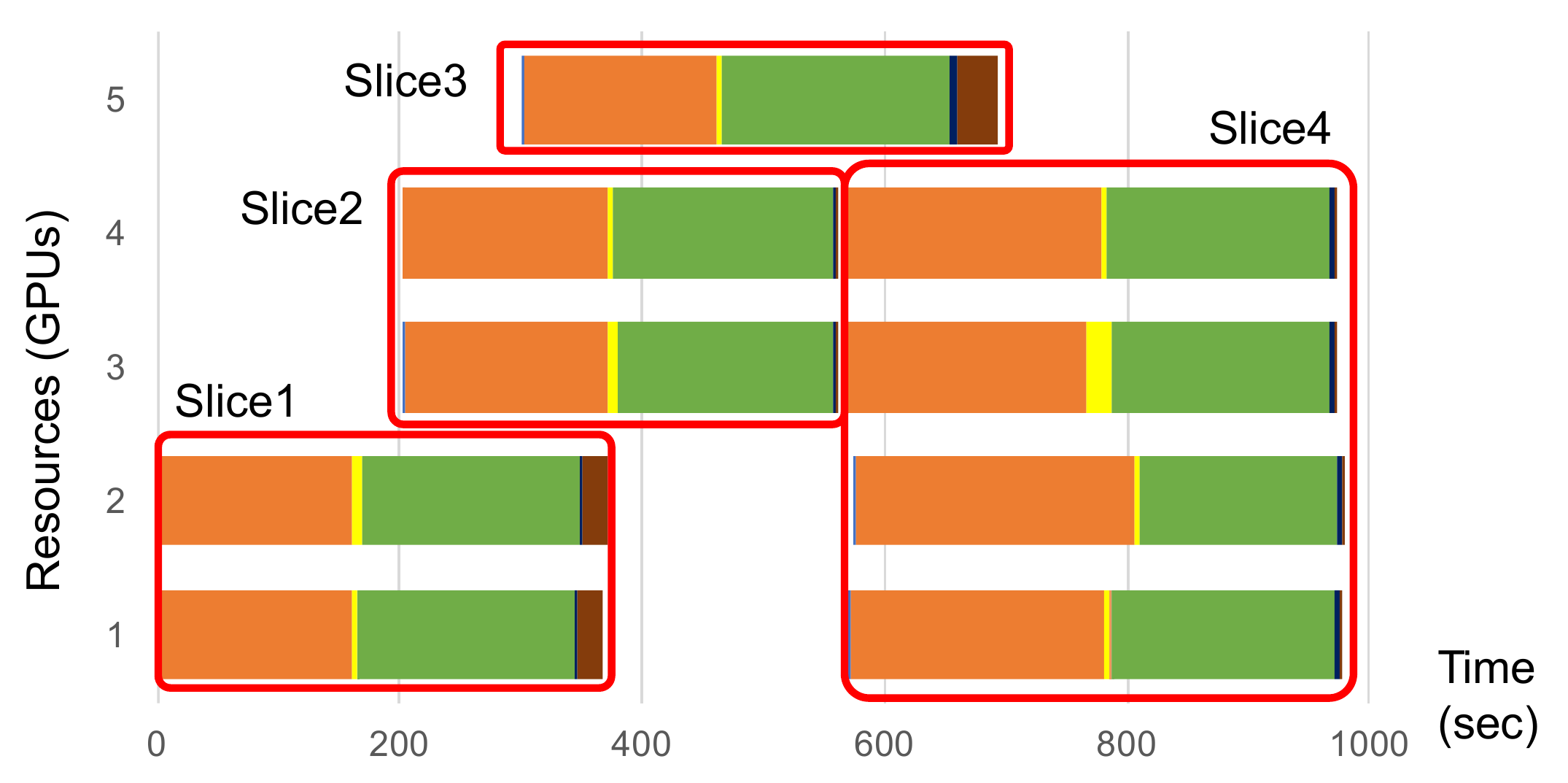}}
  \caption{Resource sharing}
  \label{fig:expr.sharing}
\end{figure}

We confirmed disaggregated resources are shared among several slices
according to a user requirement.
In this experiment, a user submitted four MNIST application jobs and FlowOS-RM allocated
resources into each slice in the FIFO manner. The slice configurations of each job are
as follows: Slice1 and 2 consist of 2node-2gpu (P100), Slice3 consists of
1node-1gpu (P40), and Slice4 consists of 4node-1gpu (P100).
Figure~\ref{fig:mnist-mix} shows that resource sharing among slices works as expected.

\section{Conclusion and Future Work}\label{sec:conclusion}

We have demonstrated flexible and effective resource sharing on the proposed
disaggregated resource management system (FlowOS-RM) for AI and Big Data applications
in next generation cloud data centers.
We found some performance issues, but the impact is limited for long, hours-running
applications like distributed deep learning training.
Our future work is replacing ExpEther with the FiC network, and then opening up
a new perspective of heterogeneous accelerator computing by leveraging resource
disaggregation.
Furthermore, in this experiment, we cannot take advantage of the potential of bare metal containers.
Thus, we plan to evaluate various applications with applying performance optimization
techniques such as a profile-guided optimization on this system.

\section*{Acknowledgement}
The authors would like to thank Hidetaka Koie, SURIGIKEN for support on the
engineering effort, and Jason Haga for his valuable comments.
This paper is partially based on results obtained from a project commissioned by
the New Energy and Industrial Technology Development Organization (NEDO).

\bibliographystyle{ieicetr}
\bibliography{flowos}

\profile{Ryousei Takano}{
is a research group leader of the Institute of Advanced Industrial Science
and Technology (AIST), Japan. He received his Ph.D. from the Tokyo University of
Agriculture and Technology in 2008. He joined AXE, Inc. in 2003 and then,
in 2008, moved to AIST. His research interests include operating systems
and distributed parallel computing. He is currently exploring an operating system
for heterogeneous accelerator clouds.}

\profile{Kuniyuasu Suzaki}{
}

\end{document}